\documentclass[12pt]{article}

\usepackage{amssymb}  
\usepackage{graphicx} 
\usepackage{psfrag}   

\def\tilde{\widetilde}

\def\nnsum#1#2{\sum_{\left\langle{#1},\,{#2}\right\rangle}}
\def\deltann#1{{\delta^{\mathrm{n.n.}}_{#1}}}
\def\tr{{\mathrm{tr}\,}}

\newcommand{\Z}{\mathbb{Z}}
\newcommand{\1}{\mbox{1\hspace{-.8ex}1}}
\newcommand{\la}{\langle}
\newcommand{\ra}{\rangle}

\begin{document}

\begin{flushright}
MPP-2005-42\\
\end{flushright}

\bigskip\bigskip

\begin{center}
{\Huge Perturbation theory for $O(3)$ topological charge 
correlators}\\
\bigskip\bigskip
{\it Miguel Aguado and Erhard Seiler \\
\bigskip
Max-Planck-Institut f\"ur Physik\\
(Werner-Heisenberg-Institut)\\
F\"ohringer Ring 6, D-80805 Munich, Germany}      
\date{\today}
\end{center}


\begin{abstract}
\noindent       
To check the consistency of positivity requirements for the two-point
correlation function of the topological charge density, which were
identified in a previous paper, we are computing perturbatively this
two-point correlation function in the two-dimensional $O(3)$ model.
We find that at the one-loop level these requirements are fulfilled.

\end{abstract}

\vskip 2mm

\section{Introduction}

In a preceding paper \cite{clash} we analyzed the implications of
physical positivity (`unitarity') and positivity of the topological
susceptibility for the two-point correlation function of the
topological density.  In particular we found that the short distance
singularity has to be softer than tree level perturbation theory would
predict.  In this article we verify for the two-dimensional $O(3)$
nonlinear sigma model that one-loop perturbation theory is indeed in
agreement with this requirement.  Such a softening of the short
distance singularity is already predicted by including the running of
the coupling constant in tree level perturbation theory \cite{bn,vic}.
But since we are dealing here with a composite operator, this check of
internal consistency of perturbation theory is quite nontrivial.

\section{Topological charge density correlator}

The action of the (ferromagnetic) O$(3)$ spin model on a 2-dimensional
square lattice $\Lambda \subseteq \mathbb{Z}^2$ is
\begin{equation}\label{action:action}
  \mathcal{S}
=
  - \beta \nnsum{X}{Y} \vec{S}_X \cdot \vec{S}_Y \ ,
\end{equation}
where the sum is over nearest neighbor pairs $\la XY \ra$ of lattice 
sites.
The partition function reads
\begin{equation}\label{action:partitionfn}
  Z
=
  \int [ \mathrm{D} S ] \,
  \mathrm{e}^{ - \mathcal{S} } ,
\end{equation}
with measure
\begin{equation}\label{action:measure}
  [ \mathrm{D} S ]
=
  \prod_x \mathrm{d}^3 {\vec{S}}_X \, \delta ( {\vec{S}}_X^{\,2} -1 ) ,
\end{equation}
imposing the unit norm constraint for all spins, i.e.  $\vec{S} \in
\mathrm{S}^2 \subset \mathbb{R}^3 \ \forall X \in \Lambda$.

We will consider the thermodynamic limit as the $L \rightarrow \infty$
limit of the model on an $L \times L$ square lattice with periodic
boundary conditions, so that translation invariance holds.  The origin
0 is then located at the center of $\Lambda$.  Unit lattice vectors
are called $\mathbf{1}$ and $\mathbf{2}$.

\subsection{Definitions of the topological charge}

We want to study the two-point function of the topological charge
density $q_X$, to be denoted by $\mathcal{G}_{\mathbf{x}}$ (with
$\mathbf{x} = \buildrel{\longrightarrow}\over{0X}$),
\begin{equation}\label{action:twopointfn}
  \mathcal{G}_{\mathbf{x}}
=
  \mathcal{G}_{0,X}
=
  \left\langle q_X q_0 \right\rangle
=
  \frac{ 1 }{ Z } \,
  \int [ \mathrm{D} S ] \,
  q_X q_0 \,
  \mathrm{e}^{ - \mathcal{S} } .
\end{equation}

The topological susceptibility is defined by
\begin{equation}\label{action:susceptibility}
  \chi_t
=
  \sum_X \mathcal{G}_{0,X} .
\end{equation}

\begin{figure}
  \centering
  \psfrag{o}{$0$}
  \psfrag{omu}{$0$+$\mathbf{1}$}
  \psfrag{omd}{$0$+$\mathbf{2}$}
  \psfrag{omt}{$0$+$\mathbf{1}$+$\mathbf{2}$}
  \psfrag{x}{$X$}
  \psfrag{xmu}{$X$+$\mathbf{1}$}
  \psfrag{xmd}{$X$+$\mathbf{2}$}
  \psfrag{xmt}{$X$+$\mathbf{1}$+$\mathbf{2}$}
  \psfrag{zu}{$Z_1$}
  \psfrag{zd}{$Z_2$}
  \psfrag{zt}{$Z_3$}
  \psfrag{zpu}{$Z'_1$}
  \psfrag{zpd}{$Z'_2$}
  \psfrag{zpt}{$Z'_3$}
  \psfrag{wu}{$W_1$}
  \psfrag{wd}{$W_2$}
  \psfrag{wt}{$W_3$}
  \psfrag{wpu}{$W'_1$}
  \psfrag{wpd}{$W'_2$}
  \psfrag{wpt}{$W'_3$}
  \includegraphics{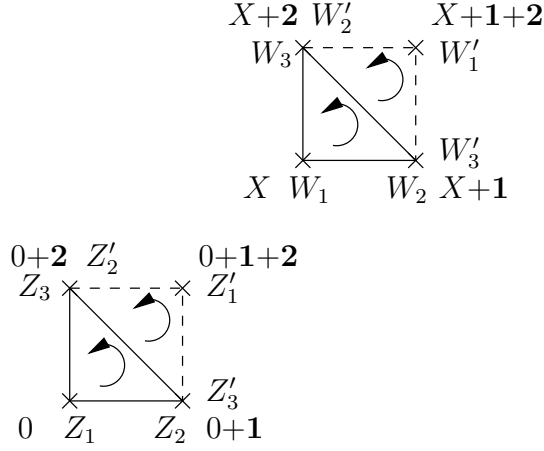}
  \caption{Triangulation of the lattice plane.}
  \label{figure:triangles}
\end{figure}

In this paper we will consider two lattice definitions of $q_X$: 
a `field theoretical' one and the geometric one due to Berg and L\"uscher. 
Both make use of a triangulation of the square lattice 
$\Z^2$ as in figure  \ref{figure:triangles}.  In both cases, the 
topological charge density is written as
\begin{equation}\label{action:densitydecomp}
  q_X
=
  q^a_X + q^b_X ,
\end{equation}
where
\begin{equation}\label{action:expressiontriangles}
{\setlength\arraycolsep{2pt}
\begin{array}{rcl}
  q^a_X
&=&
  f ( X, \, X + \mathbf{1}, \, X + \mathbf{2} ) ,
\\
&&\\
  q^b_X
&=&
  f ( X + \mathbf{1} + \mathbf{2}, \, X + \mathbf{2}, \, X + \mathbf{1} ) ,
\end{array}
}
\end{equation}
with $f : \Lambda^3 \rightarrow \mathbb{R}$ a certain function.

In \cite{clash} we used a symmetrization of this definition by
introducing in addition a `mirror' triangulation
\begin{equation}\label{action:expressionvartriangles}
{\setlength\arraycolsep{2pt}
\begin{array}{rcl}
  q^c_X
&=&
  f ( X , \, X + \mathbf{1}, \, X  + \mathbf{1} + \mathbf{2} ) ,
\\
&&\\
  q^d_X
&=&
  f ( X , \, X + \mathbf{1} + \mathbf{2}, \, X + \mathbf{2}  ) 

\end{array}
}
\end{equation}
and defining $ q_X = \frac{1}{2}( q^a_X + q^b_X + q^c_X + q^d_X) $;
this was necessary there in order to maintain reflection
symmetry. Here, however, this symmetrization is unnecessary, because
it has no effect in perturbation theory.

The 2-point function $\mathcal{G}_{\mathbf{x}} = \mathcal{G}_{0,X}$
thus has the form
\begin{equation}\label{expansion:twopointfndecomp}
  \mathcal{G}_{\mathbf{x}}
=
  \left\langle q_X q_0 \right\rangle
=
  \left\langle
    \big( q^a_X + q^b_X ) ( q^a_0 + q_0^b \big)
  \right\rangle
=
  \sum_{i,j=a,b}
  \mathcal{G}_{\mathbf{x}}^{ij}
\end{equation}
with
\begin{equation}\label{expansion:gij}
  \mathcal{G}_{\mathbf{x}}^{ij}
=
  \left\langle q^j_X q^i_0 \right\rangle ,
\qquad
  i, j = a, b .
\end{equation}

To make the notation more transparent, call
\begin{equation}\label{expansion:defzw}
{\setlength\arraycolsep{2pt}
\begin{array}{rcl}
  ( Z_1, \, Z_2, \, Z_3 )
&=&
  ( 0, \, 0 + \mathbf{1}, \, 0 + \mathbf{2} ) ,
\\
&&\\
  ( W_1, \, W_2, \, W_3 )
&=&
  ( X, \, X + \mathbf{1}, \, X + \mathbf{2} ) ,
\\
&&\\
  ( Z'_1, \, Z'_2, \, Z'_3 )
&=&
  ( 0 + \mathbf{1} + \mathbf{2}, \, 0 + \mathbf{2}, \, 0 + \mathbf{1} ) ,
\\
&&\\
  ( W'_1, \, W'_2, \, W'_3 )
&=&
  ( X + \mathbf{1} + \mathbf{2}, \, X + \mathbf{2}, \, X + \mathbf{1} )
\end{array}
}
\end{equation}
and denote $\vec{Z}_i = \vec{S}_{Z_i}$ and analogously for $W$, $Z'$,
$W'$ (see figure \ref{figure:triangles}).  The indices just introduced are
denoted by $i$, $j$, $k$, etc.

In the following sections, only formulae for $q^a_X$ and
$\mathcal{G}_{\mathbf{x}}^{aa}$ will be presented.  The full
topological charge density can be recovered by means of
(\ref{action:densitydecomp}) and (\ref{action:expressiontriangles}),
and the full two-point function by adding the other three
contributions in (\ref{expansion:twopointfndecomp}), obtained by
substituting $Z'W$, $ZW'$ and $Z'W'$ for $ZW$, respectively.
Moreover, a numerical factor will be introduced in the definition of
the correlator so as to work with simpler expressions:
\begin{equation}\label{expansion:redef}
  {\tilde{\mathcal{G}}}_{\mathbf{x}}
=
  ( 8 \pi )^2 \,
  {\mathcal{G}}_{\mathbf{x}} .
\end{equation}
%

\subsubsection{Field theoretical definition}

We first consider a symmetrized version of the field theoretical (FT)
definition \cite{ft} of the topological charge density
\begin{equation}\label{action:fieldthchargedensity}
  q^{\mathrm{FT},a}_X
=
  \frac{ 1 }{ 8 \pi } \,
  \vec{S}_X
  \cdot
  ( \vec{S}_{ X + \mathbf{1} } \wedge \vec{S}_{ X + \mathbf{2} } ) ,
\end{equation}
which gives rise to the two-point function
\begin{equation}\label{expansion:twopointfntwo}
  {\tilde{\mathcal{G}}}_{\mathbf{x}}^{\mathrm{FT},aa}
=
  \left\langle
    \left[
          \vec{Z}_1 \cdot ( \vec{Z}_2 \wedge \vec{Z}_3 )
    \right]
    \left[
          \vec{W}_1 \cdot ( \vec{W}_2 \wedge \vec{W}_3 )
    \right]
  \right\rangle
=
  \left\langle
    \det_{i,\,j} ( \vec{Z}_i \cdot \vec{W}_j )
  \right\rangle .
\end{equation}

The perturbative treatment of the problem begins with the $O(3)
\rightarrow O(2)_z$ decomposition of the spins,
\begin{equation}\label{expansion:othreetootwo}
  \vec{Z}_i
=
  \left(
        \frac{ \vec{z}_i }{ \sqrt{\beta} } \,
       , \,
        +\sqrt{ 1 - \frac{ \vec{z}_i{}^2 }{ \beta } }
  \right) ,
\qquad
  \vec{W}_j
=
  \left(
        \frac{ \vec{w}_j }{ \sqrt{\beta} } \,
       , \,
        +\sqrt{ 1 - \frac{ \vec{w}_j{}^2 }{ \beta } } \right) .
\end{equation}

Substituting in the determinant, we get
\begin{equation}\label{expansion:twopointfnyetagain}
{\setlength\arraycolsep{2pt}
\begin{array}{rcl}
  {\tilde{\mathcal{G}}}_{\mathbf{x}}^{\mathrm{FT},aa}
&=&
{\displaystyle
  \frac{ 1 }{ \beta^3 }
  \left\langle
    \det_{i,\,j} \left( \vec{z}_i \cdot \vec{w}_j \right)
  \right\rangle
}
\\
&&\\
&&\quad
{\displaystyle
 + \,
  \frac{ 1 }{ \beta^2 }
  \sum_{i,j=1}^3 (-1)^{i+j}
  \left\langle
    \sqrt{ 1 - \frac{ \vec{z}_i{}^2 }{ \beta } } \,
    \sqrt{ 1 - \frac{ \vec{w}_j{}^2 }{ \beta } } \,
    \det_{k,\ell \neq i,\,j} \left( \vec{z}_k \cdot \vec{w}_\ell \right)
  \right\rangle
}
\\
&&\\
&=&
{\displaystyle
  \frac{ 1 }{ 6 \beta^3 } \,
  \sum_{ijk\ell mn}
  \varepsilon_{ijk} \varepsilon_{\ell mn} \,
  \bigg\langle
    \vec{z}_i \cdot \vec{w}_\ell \,
    \vec{z}_j \cdot \vec{w}_m \,
    \vec{z}_k \cdot \vec{w}_n
  \bigg\rangle
}
\\
&&\\
&&\quad
{\displaystyle
 + \,
  \frac{ 1 }{ 2 \beta^2 } \,
  \sum_{ijk\ell mn}
  \varepsilon_{ijk} \varepsilon_{\ell mn} \,
  \left\langle
    \sqrt{ 1 - \frac{ \vec{z}_i{}^2 }{ \beta } } \,
    \sqrt{ 1 - \frac{ \vec{w}_\ell{}^2 }{ \beta } } \,
    \vec{z}_j \cdot \vec{w}_m \,
    \vec{z}_k \cdot \vec{w}_n
  \right\rangle .
}
\end{array}
}
\end{equation}

The perturbative expansion of (\ref{expansion:twopointfnyetagain})
consists of a Taylor expansion in $\beta^{-1/2}$. Only integral powers
of $\beta^{-1}$ contribute.  For our purposes, it will be enough to
keep terms up to and including order $\beta^{-3}$. Then
\begin{equation}\label{computation:uptothatorder}
{\setlength\arraycolsep{2pt}
\begin{array}{rcl}
  {\tilde{\mathcal{G}}}_{\mathbf{x}}^{\mathrm{FT},aa;\mathrm{pert.}}
&=&
{\displaystyle
  \frac{1}{2 \beta^2} \,
  \sum_{ijk\ell mn}
  \varepsilon_{ijk} \varepsilon_{\ell mn} \,
  \bigg\langle
    \vec{z}_j \cdot \vec{w}_m
    \vec{z}_k \cdot \vec{w}_n
  \bigg\rangle^{(0)}
}
\\
&&\\
&&
{\displaystyle
 + \,
  \frac{1}{2 \beta^3} \,
  \sum_{ijk\ell mn}
  \varepsilon_{ijk} \varepsilon_{\ell mn} \,
  \Bigg\{
    \bigg\langle
      \vec{z}_j \cdot \vec{w}_m
      \vec{z}_k \cdot \vec{w}_n
    \bigg\rangle^{(1)}
}
\\
&&\\
&&
{\displaystyle
\qquad\qquad\qquad\qquad
   - \,
    \frac{1}{6} \,
    \bigg\langle
      \big(
        3 \vec{z}_i{}^2 - 2 \vec{z}_i \cdot \vec{w}_\ell + 3 \vec{w}_\ell{}^2
      \big) \,
      \vec{z}_j \cdot \vec{w}_m
      \vec{z}_k \cdot \vec{w}_n
    \bigg\rangle^{(0)}
  \Bigg\}
}
\\
&&\\
&&
{\displaystyle
 + \,
  O ( \beta^{-4} ) ,
}
\end{array}
}
\end{equation}
where, in general, perturbative contributions to magnitude
$\mathcal{M}$ are denoted by
\begin{equation}\label{computation:pertcontrabstract}
  \mathcal{M}^{\mathrm{pert.}}
\sim
  \sum_n
  \frac{ 1 }{ \beta^n } \,
  \mathcal{M}^{(n)} .
\end{equation}
%

\subsubsection{Berg and L\"uscher's definition}

Berg and L\"uscher's (BL) definition of the topological charge density
was introduced in \cite{bl}.  With each elementary triangle 
$(X, \, Y, \, Z)$ of the chosen triangulation one associates a multiple 
of the signed area of the minimal spherical triangle determined by 
$(\vec{S}_X, \, \vec{S}_{Y}, \, \vec{S}_{Z})$ on the unit sphere.  
Explicitly,
\begin{equation}\label{bl:defqbl}
  q^{\mathrm{BL},a}_X
=
  \frac{ 1 }{ 2 \pi } \,
  \tan^{-1}
  \frac{
        \vec{S}_X \cdot ( \vec{S}_{X+\mathbf{1}} \wedge \vec{S}_{X+\mathbf{2}} )
      }{
        1
        +
        \vec{S}_X \cdot \vec{S}_{X+\mathbf{1}}
        +
        \vec{S}_{X+\mathbf{1}} \cdot \vec{S}_{X+\mathbf{2}}
        +
        \vec{S}_{X+\mathbf{2}} \cdot \vec{S}_X
       } .
\end{equation}

The corresponding perturbative two-point function is, up to and
including order $\beta^{-3}$,
\begin{equation}\label{bl:twopoint}
{\setlength\arraycolsep{0pt}
\begin{array}{rl}
&
  {\tilde{\mathcal{G}}}^{\mathrm{BL},aa;\mathrm{pert.}}_X
\\
&\\
&=
{\displaystyle
  16 \,
  \left\langle
  \frac{
        \det_{i,j} ( \vec{Z}_i \cdot \vec{W}_j )
      }{
        \big(
             1
             +
             \vec{Z}_1 \cdot \vec{Z}_2
             + 
             \vec{Z}_2 \cdot \vec{Z}_3 
             +
             \vec{Z}_3 \cdot \vec{Z}_1
        \big)
        \big(
             1
             +
             \vec{W}_1 \cdot \vec{W}_2
             +
             \vec{W}_2 \cdot \vec{W}_3
             +
             \vec{W}_3 \cdot \vec{W}_1
        \big)
       }
  \right\rangle
}
\\
&\\
&
\quad
 + \,
  O ( \beta^{-4} ) .
\end{array}
}
\end{equation}

Substituting (\ref{expansion:othreetootwo}) in (\ref{bl:twopoint}) and
Taylor expanding the result, we arrive at
\begin{equation}\label{bl:twopointftvsbl}
  {\tilde{\mathcal{G}}}_{\mathbf{x}}^{\mathrm{BL},aa;\mathrm{pert.}}
=
  {\tilde{\mathcal{G}}}_{\mathbf{x}}^{\mathrm{FT},aa;\mathrm{pert.}}
 +
  \Delta {\tilde{\mathcal{G}}}^{aa}_{\mathbf{x}} ,
\end{equation}
where
\begin{equation}\label{bl:bldiff}
{\setlength\arraycolsep{2pt}
\begin{array}{rcl}
  \Delta {\tilde{\mathcal{G}}}^{aa}_{\mathbf{x}}
&=&
{\displaystyle
  \frac{ 1 }{ 8 \beta^3 } \,
  \sum_{ijk\ell mn}
  \varepsilon_{ijk} \varepsilon_{\ell mn}
  \left\langle
    \left(
      \sum_{u=1}^3
        ( \vec{z}_u{}^2 + \vec{w}_u{}^2 )
     -
      \sum_{u < v}
        ( \vec{z}_u \cdot \vec{z}_v + \vec{w}_u \cdot \vec{w}_v )
    \right)
  \right\rangle^{(0)}
}
\\
&&\\
&&
{\displaystyle
 + \,
  O ( \beta^{-4} ) .
}
\end{array}
}
\end{equation}
will be referred to as `BL difference.'

\subsection{Perturbative computation of ${\tilde{\mathcal{G}}}_x$}

In this section we use the BL topological charge density
(\ref{bl:defqbl}) and will comment on the difference with the FT
definition when appropriate.  Therefore we drop the BL labels and
write
\begin{equation}\label{treelevel:gpert}
  {\tilde{\mathcal{G}}}_{\mathbf{x}}^{\mathrm{pert.}}
=
  \frac{ 1 }{ \beta^2 } \,
  {\tilde{\mathcal{G}}}_{\mathbf{x}}^{\mathrm{T}}
 +
  \frac{ 1 }{ \beta^3 } \,
  {\tilde{\mathcal{G}}}_{\mathbf{x}}^{\mathrm{L}}
 +
  O ( \beta^{-4} ) .
\end{equation}

For bookkeeping purposes we do the perturbative computation of the density 
correlator out keeping factors $N-1$ explicit, as for an O$(N)$ problem, 
although the topological density has a meaning only for $N=3$.

The only vertices relevant for Feynman diagrams up to and including
order $\beta^{-3}$ are order $\beta^{-1}$:
\begin{itemize}
\item[$\bullet$] Two-$\vec{\pi}$ vertex:
\begin{equation}\label{computation:twopivertex}
  V_{ZW}^{k\ell}
=
  \frac{1}{\beta} \,
  \left( 1 - \frac{N-1}{V} \right)
  \delta^{k\ell} \delta_{ZW} .
\end{equation}

\item[$\bullet$] Four-$\vec{\pi}$ vertex:
\begin{equation}\label{computation:fourpivertex}
{\setlength\arraycolsep{2pt}
\begin{array}{rl}
&
{\displaystyle
  V_{ZWTU}^{k\ell mn}
=
  \frac{1}{\beta} \,
  \left[
        - n_Z \delta_{ZWTU}
         \Big(
              \delta^{k\ell} \delta^{mn}
              + \delta^{km} \delta^{\ell n}
              + \delta^{kn} \delta^{\ell m}
         \Big)
  \right.
}
\\
&\\
&\quad
{\displaystyle
  \left.
        + \Big(
                \deltann{ZT} \delta_{ZW} \delta_{TU}
                \delta^{k\ell} \delta^{mn}
               +
                \deltann{ZW} \delta_{ZT} \delta_{WU}
                \delta^{km} \delta^{\ell n}
               +
                \deltann{ZW} \delta_{ZU} \delta_{WT}
                \delta^{kn} \delta^{\ell m}
          \Big)
  \right] ,
}
\end{array}
}
\end{equation}
\end{itemize}
where subindices are lattice points, superindices are O$(N)$ indices,
$n_Z = 4$ is the number of nearest neighbors of point $Z$, two-index
deltas are Kronecker,
\begin{equation}\label{treelevel:defdeltazwtu}
  \delta_{ZWTU}
=
  \left\{
    \begin{array}{lcl}
      1 && \mathrm{if}\ Z=W=T=U , \\
      0 && \mathrm{otherwise}
    \end{array}
  \right.
\end{equation}
and
\begin{equation}\label{treelevel:defdeltann}
  \deltann{ZW}
=
  \left\{
    \begin{array}{lcl}
      1 && \mathrm{if}\ Z, \, W\ \mathrm{nearest\ neighbors}, \\
      0 && \mathrm{otherwise} .
    \end{array}
  \right.
\end{equation}

The explicitly volume-dependent factor in
(\ref{computation:twopivertex}) comes from fixing one spin using the
Faddeev-Popov method as in \cite{has} to eliminate the divergent zero
mode that would otherwise appear in the Gaussian measure defining tree
level perturbation theory with periodic boundary conditions.

\subsubsection{Green's functions}
\label{sec:green}

In order to compute the Green's functions relevant to the correlators
being studied, we start from the corresponding quantity on the $L
\times L$ square lattice and put $\mathbf{x} =
\buildrel{\longrightarrow}\over{0X} = ( x, \, y )$, $-L/2 \leq x, y <
L/2$.
\begin{equation}\label{green:lattice}
  G_{0,X}
=
  \frac{ a^2 }{ 2 L^2 } \,
  \sum_{ \mathbf{p} }
  {\vphantom\sum}'
  \frac{
        \cos ( a \, \mathbf{p} \cdot \mathbf{x} )
      }{
        2 - \cos ( a \, p_x ) - \cos ( a \, p_y )
       } \, ,
\end{equation}
where the momenta
\begin{equation}\label{green:momentumlattice}
  \mathbf{p}
=
  ( p_x, \, p_y )
=
  \frac{ 2 \pi }{ L } \, ( n_x, \, n_y ) ,
\quad
  n_x, n_y \in \mathbb{Z}
\end{equation}
are summed over the first Brillouin zone $-\, L/2 \leq n_x, n_y < L/2$,
the prime meaning exclusion of the zero mode $\mathbf{p} = \mathbf{0}$.

The thermodynamic ($L \rightarrow \infty$) limit of
(\ref{green:lattice}) does not exist, since the sum diverges
logarithmically in $L$. The divergence is independent of $\mathbf{x}$,
however, and the thermodynamic limit of the difference $\tilde
G_{0,X}=G_{0,X}-G_\mathbf{0}$ does exist. According to a general
argument due to David \cite{david}, the perturbation expansion of
invariant observables should be free of infrared divergences; the
first sign of this is the fact that everything is expressed in terms
of $\tilde G_{0,X}$, which has a well defined thermodynamic limit
given by the integral
\begin{equation}\label{green:integralsubtracted}
  {\tilde{G}}_{0,X}
\equiv
  G_{0,X} - G_{ \mathbf{0} }
=
  \frac{ 1 }{ 2 ( 2 \pi )^2 } \,
  \int_{-\pi}^{+\pi}
  \mathrm{d} q_x
  \int_{-\pi}^{+\pi}
  \mathrm{d} q_y \,
  \frac{
        \cos ( \mathbf{q} \cdot \mathbf{x} ) - 1
      }{
        2 - \cos q_x - \cos q_y
       }\ ; 
\end{equation}
(another possible source of infrared divergence is the sum over
intermediate lattice sites, which will be addressed in section
\ref{section:numeval}). Obviously ${\tilde{G}}_{\mathbf{0}} = 0$,
${\tilde{G}}_{( x, \, y )} = {\tilde{G}}_{( y, \, x )}$.

This subtracted propagator was studied by other methods in
\cite{shin}, which we will use as a check for our results.

Due to parity, the numerator in the integrand of
(\ref{green:integralsubtracted}) can be replaced by $\cos ( x q_x )
\cos ( y q_y ) - 1$.  Integration in $q_x$ can be performed by the
calculus of residues, with the result
\begin{equation}\label{green:integralchebyshev}
  {\tilde{G}}_{0,X}
=
  \frac{ 1 }{ 2 \pi } \,
  \int_{-1}^{+1}
  \frac{
        \mathrm{d} t
      }{
        \sqrt{ 1 - t^2 }
       } \,
  \frac{
        \big( 2 - t + \sqrt{ 3 - 4 t + t^2 } \big)^{-x}
        \mathrm{T}_y (t)
        -
        1
      }{
        \sqrt{ 3 - 4 t + t^2 }
       } .
\end{equation}
Here T$_y (t)$ is Chebyshev's polynomial of the $y$-th kind.

It is straightforward to check that (\ref{green:integralchebyshev})
behaves asymptotically as
\begin{equation}\label{green:asymptotics}
  {\tilde{G}}_{0,X}
=
 - \,
  \frac{ 1 }{ 2 \pi } \,
  \ln |\mathbf{x}|  
 - \,
  c
 - \,
  \frac{ d }{ |\mathbf{x}| }
 + \,
  O ( |\mathbf{x}|^{-2} )
\end{equation}
with $c$, $d$ constants. According to \cite{shin},
\begin{equation}\label{green:valuesofcd}
  c = \frac{ 2 \gamma_E + 3 \ln 2 }{ 4 \pi } \, ,
\qquad
  d = 0 ,
\end{equation}
where $\gamma_E$ is Euler's constant.

Similarly it can be seen that 
\begin{equation}\label{2ndasymp}
  {\tilde{G}}_{0,X+\mathbf{v}}+{\tilde{G}}_{0,X-\mathbf{v} }-
 2{\tilde{G}}_{0,X}
=
  \frac{ 1 }{ \pi } \, 
  \left(
     \frac{ \mathbf{v}^2 }{ 2 \, \mathbf{x}^2 }
    - \,
     \frac{ ( \mathbf{v} \cdot \mathbf{x} )^2 }{ |\mathbf{x}|^4 }
  \right)
 + \,
  O ( |\mathbf{x}|^{-3} )\ 
\end{equation}
and
\begin{equation}\label{mixasymp}
 {\tilde{G}}_{0,X+\mathbf{v}}-{\tilde{G}}_{0,X+\mathbf{w} }+
 {\tilde{G}}_{0,X-\mathbf{v}}-{\tilde{G}}_{0,X+\mathbf{w} }
=
  \frac{ 1 }{ 2\pi } \,
  \frac{ \mathbf{v}^2-\mathbf{w}^2- 
        2 ( \mathbf{v} \cdot \mathbf{x} )^2 +
        2 ( \mathbf{w} \cdot \mathbf{x} )^2}
   { |\mathbf{x}|^4 }
 + \,
  O ( |\mathbf{x}|^{-3} )\  
\end{equation}
for large $|\mathbf{x}|$ and $|\mathbf{v}|, |\mathbf{w}| = O(1)$.  To
see this, one subtracts from the Fourier representation of the lhs of
(\ref{2ndasymp}) and \ref{mixasymp} the corresponding continuum
expression with a suitable smooth high momentum cutoff; this
difference, being the Fourier transform of a smooth (${\cal
C}^\infty$) function will decay faster than any power of
$|\mathbf{x}|$.  The cutoff on the continuum expression, on the other
hand, does not affect the leading long distance behavior, which is
obtained straightforwardly.  Expressions (\ref{2ndasymp}) and
(\ref{mixasymp}), in which the constants $c$ and $d$ in
(\ref{green:asymptotics}) do not appear, will be useful in the
computation of the charge correlator.

In the following, we will denote Green's functions by $G_{0,X}$ even
when subtracted.

\subsubsection{Tree level}

\begin{figure}
  \centering
  \includegraphics{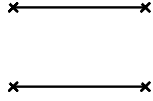}
  \caption{Tree level diagram structure.}
  \label{graph:treelevel}
\end{figure}

The tree level contribution to
${\tilde{\mathcal{G}}}_{\mathbf{x}}^{11}$ is given by
\begin{equation}\label{treelevel:gtwo}
  {\tilde{\mathcal{G}}}_{\mathbf{x}}^{\mathrm{T},aa}
=
  \frac{1}{2} \,
  \sum_{ijk\ell mn}
  \varepsilon_{ijk} \varepsilon_{\ell mn}
  \bigg\langle
    \vec{z}_j \cdot \vec{w}_m
    \vec{z}_k \cdot \vec{w}_n
  \bigg\rangle^{(0)} ,
\end{equation}
the correlator in the rhs being evaluated in the Gaussian measure with
one spin fixed.  The corresponding diagrams have the structure of
figure (\ref{graph:treelevel}).

Contracting spin indices,
\begin{equation}\label{treelevel:afterdiagrams}
{\setlength\arraycolsep{2pt}
\begin{array}{rcl}
  {\tilde{\mathcal{G}}}_{\mathbf{x}}^{\mathrm{T},aa}
&=&
{\displaystyle
  \frac{1}{2} \,
  (N-1)^2 \,
  \sum_{ijk\ell mn}
  \varepsilon_{ijk} \varepsilon_{\ell mn}
  G_{ Z_j W_m } G_{ Z_k W_n }
}
\\
&&\\
&&
{\displaystyle
 + \,
  \frac{1}{2} \,
  (N-1) \,
  \sum_{ijk\ell mn}
  \varepsilon_{ijk} \varepsilon_{\ell mn}
  G_{ Z_j Z_k } G_{ W_m W_n }
}
\\
&&\\
&&
{\displaystyle
 + \,
  \frac{1}{2} \,
  (N-1) \,
  \sum_{ijk\ell mn}
  \varepsilon_{ijk} \varepsilon_{\ell mn}
  G_{ Z_j W_n } G_{ Z_k W_m } .
}
\end{array}
}
\end{equation}

The $3 \times 3$ matrices $G_{ Z_\cdot Z_\cdot }$ and $G_{ W_\cdot
W_\cdot }$ are symmetric, and therefore the second term in the rhs
vanishes.  Playing with indices, we can recast equation
(\ref{treelevel:afterdiagrams}) in the form
\begin{equation}\label{treelevel:afterdiagramsagain}
  {\tilde{\mathcal{G}}}_{\mathbf{x}}^{\mathrm{T},aa}
=
  \frac{1}{2} \,
  \left[ (N-1)^2 - (N-1) \right]
  \sum_{ijk\ell mn}
  \varepsilon_{ijk} \varepsilon_{\ell mn}
  G_{ Z_j W_m } G_{ Z_k W_n } ,
\end{equation}

This can be written in terms of differences $G_{\mathbf{v}} -
G_{\mathbf{w}}$ with $|\mathbf{v}- \mathbf{w}| \leq \sqrt{2}$ (which
restricts IR divergences to summation effects, the IR divergences of
Green's functions being cancelled in the differences):
\begin{equation}\label{treelevel:afterdiagramsagaindiff}
{\setlength\arraycolsep{2pt}
\begin{array}{rcl}
  {\tilde{\mathcal{G}}}_{\mathbf{x}}^{\mathrm{T},aa}
&=&
{\displaystyle
  - \,
  \frac{1}{2} \,
  \left[ (N-1)^2 - (N-1) \right]
}
\\
&&\\
&&
{\displaystyle
\qquad
\times
  \sum_{ijk\ell mn}
  \varepsilon_{ijk} \varepsilon_{\ell mn}
  ( G_{ Z_j W_m } - G_{ Z_k W_m } )
  ( G_{ Z_k W_m } - G_{ Z_k W_n } )
}
\end{array}
}
\end{equation}

We can now perform the sums over all indices: in general, for a $3
\times 3$ matrix $a_{ij}$,
\begin{equation}\label{treelevel:sumsoverindices}
{\setlength\arraycolsep{2pt}
\begin{array}{rcl}
{\displaystyle
  \sum_{ijk\ell mn}
  \varepsilon_{ijk} \varepsilon_{\ell mn}
  a_{jm} a_{kn}
}
&=&
{\displaystyle
  \sum_{i\ell}
  \left\{
    \left[ (\tr a)^2 - \tr(a^2) \right] \delta_{\ell i}
   +
    2 ( a^2 )_{\ell i}
   -
    2 ( \tr a ) a_{\ell i}
  \right\}
}
\\
&&\\
&=&
{\displaystyle
  3 \left[ (\tr a)^2 - \tr(a^2) \right]
 +
  2 \, \mathrm{s}( a^2 )
 -
  2 ( \tr a ) \, \mathrm{s}(a) ,
}
\end{array}
}
\end{equation}
where $\mathrm{s}(a)$ denotes the sum of all entries in $a$.

This identity can be obtained as follows: one 
starts from the well-known identity  
\begin{equation}
\ln\det(\1+zA)= \tr \ln (\1+zA)
\end{equation}
and expands it to order $z^3$ to obtain
\begin{equation}
\tr A\wedge A\wedge A=\frac{1}{3} \tr A^3- \frac{1}{2} \tr A\ \  \tr A^2+ 
\frac {1}{6} (\tr A)^3\ .
 \end{equation}
By `polarization', i.e. the replacement of $A$ by $xA+yB+zC$ on both 
sides and comparing the coefficient of $xyz$  
Eq.~(\ref{treelevel:sumsoverindices}) follows if we put $B=C=a$ 
and take for $A$ the matrix with all entries equal to 1.

In our case,
\begin{equation}\label{treelevel:finalexprgtree}
{\setlength\arraycolsep{2pt}
\begin{array}{rcl}
  {\tilde{\mathcal{G}}}_{\mathbf{x}}^{\mathrm{T},aa}
&=&
{\displaystyle
  - \,
  \frac{1}{2} \,
  \left[ (N-1)^2 - (N-1) \right]
}
\\
&&\\
&&
{\displaystyle
\quad
 \times
  \left\{
    3 \left[ (\tr G_{ Z_\cdot W_\cdot } )^2 - \tr( G_{ Z_\cdot W_\cdot }^2 ) \right]
   +
    2 \, \mathrm{s}( G_{ Z_\cdot W_\cdot }^2 )
   -
    2 ( \tr G_{ Z_\cdot W_\cdot } ) \, \mathrm{s}( G_{ Z_\cdot W_\cdot } )
  \right\} ,
}
\end{array}
}
\end{equation}
%

\subsubsection{One loop}

\begin{figure}
  \centering
  \includegraphics{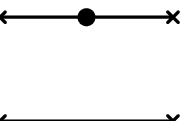} \hspace{1cm}%
  \includegraphics{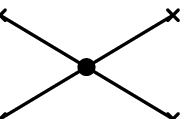} \hspace{1cm}%
  \includegraphics{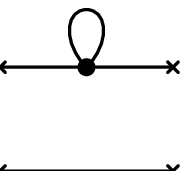} \hspace{1cm}%
  \includegraphics{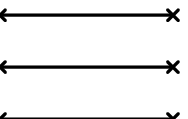}
  \caption{One-loop diagram structures.}
  \label{graph:loop}
\end{figure}

The one-loop (order $\beta^{-3}$) correction to the correlator can be
decomposed as
\begin{equation}\label{thegraphs:firstcorrection}
  {\tilde{\mathcal{G}}}_{\mathbf{x}}^{\mathrm{L},aa}
=
  (\mathrm{I})^{aa} + (\mathrm{II})^{aa} + (\mathrm{III})^{aa}
  + (\mathrm{IV})^{aa} + (\mathrm{V})^{aa} ,
\end{equation}
according to the structures depicted in figure (\ref{graph:loop}).

$\bullet$ (I) contains the contributions of the 4-legged, disconnected
graphs with one 2-vertex:
\begin{equation}\label{thegraphs:graphi}
{\setlength\arraycolsep{2pt}
\begin{array}{rcl}
  (\mathrm{I})^{aa}
&=&
{\displaystyle
  \left( 1 - \frac{N-1}{V} \right)
  \left[ (N-1)^2 - (N-1) \right]
}
\\
&&\\
&&
{\displaystyle
\qquad
\times
  \sum_{ijk\ell mn}
  \varepsilon_{ijk} \varepsilon_{\ell mn}
  G_{Z_j W_m}
  \sum_P
    G_{Z_k P} G_{W_n P} .
}
\end{array}
}
\end{equation}

$\bullet$ (II) contains the X-shaped contributions:
\begin{equation}\label{thegraphs:graphii}
{\setlength\arraycolsep{2pt}
\begin{array}{rcl}
  (\mathrm{II})^{aa}
&=&
{\displaystyle
  - \,
  \frac{1}{2} \,
  \left[ (N-1)^2 - (N-1) \right]
  \sum_{ijk\ell mn}
  \varepsilon_{ijk} \varepsilon_{\ell mn}
}
\\
&&\\
&&\qquad
{\displaystyle
\times
  \nnsum{P}{Q}
  \left( G_{Z_j P} G_{W_m P} - G_{Z_j Q} G_{W_m Q} \right)
  \left( G_{Z_k P} G_{W_n P} - G_{Z_k Q} G_{W_n Q} \right) .
}
\end{array}
}
\end{equation}

$\bullet$ (III) gathers the contributions of 4-legged graphs with
tadpoles:
\begin{equation}\label{thegraphs:graphiii}
{\setlength\arraycolsep{2pt}
\begin{array}{rcl}
  (\mathrm{III})^{aa}
&=&
{\displaystyle
  - \,
  \frac{N-1}{2} \,
  \left[ (N-1)^2 - (N-1) \right]
  \sum_{ijk\ell mn}
  \varepsilon_{ijk} \varepsilon_{\ell mn}
  G_{Z_j W_m}
}
\\
&&\\
&&
{\displaystyle
\qquad\qquad
 \times
  \nnsum{P}{Q}
  ( G_{PP} - G_{QQ} )
  \left( G_{Z_k P} G_{W_n P} - G_{Z_k Q} G_{W_n Q} \right)
}
\\
&&\\
&&
{\displaystyle
  - \,
  \left[ (N-1)^2 - (N-1) \right]
  \sum_{ijk\ell mn}
  \varepsilon_{ijk} \varepsilon_{\ell mn}
  G_{Z_j W_m}
}
\\
&&\\
&&
{\displaystyle
\qquad\qquad
 \times
  \nnsum{P}{Q}
  \big[
        G_{PP} G_{Z_k P} G_{W_n P} + G_{QQ} G_{Z_k Q} G_{W_n Q}
}
\\
&&\\
&&
{\displaystyle
\qquad\qquad\qquad\qquad
       -
        G_{PQ} \left( G_{Z_k P} G_{W_n Q} + G_{Z_k Q} G_{W_n P} \right)
  \big] .
}
\end{array}
}
\end{equation}
The first term of (III) vanishes by translation invariance.

$\bullet$ (IV) consists of the six-legged graphs contributing to the
correlator according to the FT definition of the charge density:
\begin{equation}\label{thegraphs:graphiv}
{\setlength\arraycolsep{2pt}
\begin{array}{rcl}
  (\mathrm{IV})^{aa}
&=&
{\displaystyle
  \frac{N-3}{6} \,
  \left[ (N-1)^2 - (N-1) \right]
  \det G_{Z_\cdot W_\cdot}
}
\\
&&\\
&&
{\displaystyle
  - \,
  \frac{N-1}{2} \,
  \left[ (N-1)^2 - (N-1) \right]
  \sum_{ijk\ell mn}
  \varepsilon_{ijk} \varepsilon_{\ell mn}
  G_{00} G_{Z_j W_m} G_{Z_k W_n}
}
\\
&&\\
&&
{\displaystyle
  - \,
  \left[ (N-1)^2 - (N-1) \right]
}
\\
&&\\
&&
{\displaystyle
\qquad
 \times
  \sum_{ijk\ell mn}
  \varepsilon_{ijk} \varepsilon_{\ell mn}
  G_{Z_j W_m}
  \left(
        G_{Z_i Z_k} G_{Z_i W_n} + G_{Z_k W_\ell} G_{W_\ell W_n}
  \right) .
}
\end{array}
}
\end{equation}
The determinant term in (IV)$^{aa}$ has a vanishing prefactor.

$\bullet$ (V) is the BL difference, also consisting of six-legged
graphs:
\begin{equation}\label{thegraphs:graphv}
{\setlength\arraycolsep{2pt}
\begin{array}{rcl}
  (\mathrm{V})^{aa}
&=&
{\displaystyle
  - \,
  \frac{N-1}{8} \,
  \left[ (N-1)^2 - (N-1) \right]
  \left[ \sum_{u,v} \left( G_{Z_u Z_v} - G_{00} \right) \right]
}
\\
&&\\
&&
{\displaystyle
\qquad
 \times
  \sum_{ijk\ell mn}
  \varepsilon_{ijk} \varepsilon_{\ell mn}
  G_{Z_j W_m} G_{Z_k W_n}
}
\\
&&\\
&&
{\displaystyle
  - \,
  \frac{1}{4} \,
  \left[ (N-1)^2 - (N-1) \right]
  \sum_{ijk\ell mn}
  \varepsilon_{ijk} \varepsilon_{\ell mn}
  G_{Z_j W_m}
}
\\
&&\\
&&
{\displaystyle
\qquad
 \times
  \sum_{u,v}
  \left[
        G_{Z_u Z_k} \left( G_{Z_v W_n} - G_{Z_u W_n} \right)
       +
        G_{Z_k W_u} \left( G_{W_n W_v} - G_{W_n W_u} \right)
  \right] .
}
\end{array}
}
\end{equation}

As happened with the tree level expression, the one-loop correction
can be rewritten in terms of differences $G_{\mathbf{v}} -
G_{\mathbf{w}}$ with $|\mathbf{v}- \mathbf{w}| \leq \sqrt{2}$.  This
works independently for (I), (II) and (V):
\begin{equation}\label{thegraphs:graphidiff}
{\setlength\arraycolsep{2pt}
\begin{array}{rcl}
  (\mathrm{I}_d)^{aa}
&=&
{\displaystyle
  \frac{ 1 }{ 3 } \,
  \left( 1 - \frac{N-1}{V} \right)
  \left[ (N-1)^2 - (N-1) \right]
  \sum_{ijk\ell mn}
  \varepsilon_{ijk} \varepsilon_{\ell mn}
}
\\
&&\\
&&
{\displaystyle
\qquad
\times
  ( G_{Z_j W_m} - G_{Z_k W_m} )
  \sum_P
    ( G_{Z_k P} - G_{Z_i P} )
    ( G_{W_n P} - G_{W_m P} ) ,
}
\end{array}
}
\end{equation}
%


%
\begin{equation}\label{thegraphs:graphiidiff}
{\setlength\arraycolsep{2pt}
\begin{array}{rl}
&
{\displaystyle
  (\mathrm{II}_d)^{aa}
=
  \frac{1}{16}
  \left[ (N-1)^2 - (N-1) \right]
  \sum_{ijk\ell mn}
  \varepsilon_{ijk} \varepsilon_{\ell mn}
}
\\
&\\
&
{\displaystyle
\times
  \nnsum{P}{Q}
  \big[ ( G_{Z_j P} - G_{Z_j Q} ) - ( G_{W_m P} - G_{W_m Q} ) \big]
  \big[ ( G_{Z_k P} - G_{Z_k Q} ) - ( G_{W_n P} - G_{W_n Q} ) \big]
}
\\
&\\
&\quad
{\displaystyle
\times
  \big[
        ( G_{Z_j P} - G_{Z_k P} ) + ( G_{Z_j Q} - G_{Z_k Q} )
      - ( G_{W_m P} - G_{W_n P} ) - ( G_{W_m Q} - G_{W_n Q} )
  \big]^2 ,
}
\end{array}
}
\end{equation}
%


%
\begin{equation}\label{thegraphs:graphvdiff}
{\setlength\arraycolsep{2pt}
\begin{array}{rcl}
  (\mathrm{V_d})^{aa}
&=&
{\displaystyle
  \frac{N-1}{8} \,
  \left[ (N-1)^2 - (N-1) \right]
  \left[ \sum_{u,v} \left( G_{Z_u Z_v} - G_{00} \right) \right]
}
\\
&&\\
&&\qquad
{\displaystyle
\times
  \sum_{ijk\ell mn}
  \varepsilon_{ijk} \varepsilon_{\ell mn}
  ( G_{Z_j W_m} - G_{Z_k W_m} )
  ( G_{Z_k W_m} - G_{Z_k W_n} )
}
\\
&&\\
&&
{\displaystyle
  - \,
  \frac{1}{8} \,
  \left[ (N-1)^2 - (N-1) \right]
  \sum_{ijk\ell mn}
  \varepsilon_{ijk} \varepsilon_{\ell mn}
  ( G_{Z_j W_m} - G_{Z_k W_m} )
}
\\
&&\\
&&\qquad
{\displaystyle
\times
  \sum_{u,v}
  \big[
       ( G_{Z_u Z_k} - G_{Z_v Z_k} ) ( G_{Z_v W_n} - G_{Z_u W_n} )
}
\\
&&\\
&&\qquad\qquad\quad
{\displaystyle
 +
        ( G_{Z_k W_u} - G_{Z_k W_v} ) ( G_{W_n W_v} - G_{W_n W_u} )
  \big] ,
}
\end{array}
}
\end{equation}
while (III)$^{aa}$ and (IV)$^{aa}$ cannot be brought to that form.
However, the sum (III)$^{aa}$ + (IV)$^{aa}$ can, because, using
translation invariance and the fact that $N = 3$,
\begin{equation}\label{thegraphs:graphiiidiff}
{\setlength\arraycolsep{2pt}
\begin{array}{rcl}
  (\mathrm{III}_d)^{aa}
&=&
{\displaystyle
  2 G_{00}
  \sum_{ijk\ell mn}
  \varepsilon_{ijk} \varepsilon_{\ell mn}
  ( G_{Z_j W_m} - G_{Z_k W_m} )
  ( G_{Z_k W_m} - G_{Z_j W_n} )
}
\\
&&\\
&&
{\displaystyle
  - \,
  2
  ( G_{01} - G_{00} )
  \sum_{ijk\ell mn}
  \varepsilon_{ijk} \varepsilon_{\ell mn}
  ( G_{Z_j W_m} - G_{Z_k W_m} )
}
\\
&&\\
&&\qquad
{\displaystyle
\times
  \nnsum{P}{Q}
  ( G_{Z_k P} - G_{Z_k Q} ) ( G_{W_n P} - G_{W_n Q} )
}
\\
&&\\
&&
{\displaystyle
  + \,
  \frac{8}{3}
  ( G_{01} - G_{00} )
  \sum_{ijk\ell mn}
  \varepsilon_{ijk} \varepsilon_{\ell mn}
  ( G_{Z_j W_m} - G_{Z_k W_m} )
}
\\
&&\\
&&\qquad
{\displaystyle
\times
  \nnsum{P}{Q}
  ( G_{Z_k P} - G_{Z_i P} ) ( G_{W_n P} - G_{W_m P} )
}
\end{array}
}
\end{equation}
%
and
\begin{equation}\label{thegraphs:graphivdiff}
{\setlength\arraycolsep{2pt}
\begin{array}{rcl}
  (\mathrm{IV}_d)^{aa}
&=&
{\displaystyle
  - \,
  2 G_{00}
  \sum_{ijk\ell mn}
  \varepsilon_{ijk} \varepsilon_{\ell mn}
  ( G_{Z_j W_m} - G_{Z_k W_m} )
  ( G_{Z_k W_m} - G_{Z_j W_n} )
}
\\
&&\\
&&
{\displaystyle
  - \,
  \sum_{ijk\ell mn}
  \varepsilon_{ijk} \varepsilon_{\ell mn}
  ( G_{Z_i Z_k} - G_{00} )
  ( G_{Z_j W_m} - G_{Z_j W_n} )
  ( G_{Z_i W_n} - G_{Z_k W_n} )
}
\\
&&\\
&&
{\displaystyle
  - \,
  \sum_{ijk\ell mn}
  \varepsilon_{ijk} \varepsilon_{\ell mn}
  ( G_{W_\ell W_n} - G_{00} )
  ( G_{Z_j W_m} - G_{Z_k W_m} )
  ( G_{Z_j W_\ell} - G_{Z_k W_n} ) .
}
\end{array}
}
\end{equation}
The first terms in the rhs of (\ref{thegraphs:graphiiidiff}) and
(\ref{thegraphs:graphivdiff}) cancel.

\subsection{$\zeta_{ab}$ notation}

Green's functions appearing (up to 1-loop order) in the perturbative
expansion of ${\tilde{\mathcal{G}}}_{\mathbf{x}}$ are of the form
$G_{\mathbf{v}+\mathbf{w}}$, where
\begin{itemize}
\item $\mathbf{v}$ is one of $\mathbf{\zeta} \equiv
\buildrel{\longrightarrow}\over{0X}$, $\mathbf{\xi} \equiv
\buildrel{\longrightarrow}\over{P0}$ (where $P$ is summed over all
lattice points), $\mathbf{\eta} \equiv
\buildrel{\longrightarrow}\over{PX}$ and $\mathbf{\kappa} \equiv
\mathbf{0} \equiv \buildrel{\longrightarrow}\over{00}$,
\item $\mathbf{w}$ is a linear combination $n_1 \mathbf{1} + n_2
\mathbf{2}$ of unit lattice vectors with $n_i$ integers ($|n_i| \leq
2$ up to 1-loop order.)
\end{itemize}

The following notation proves useful when applying computer techniques
to this perturbative problem:
\begin{equation}\label{notation:recipe}
  G_{\mathbf{v}+\mathbf{w}}
\rightarrow
  v_{ \mathrm{word} }.
\end{equation}
Here $v$ is one of the {\sl letters}\/ $\zeta$, $\xi$, $\eta$ or
$\kappa$ according to the vector $\mathbf{v}$.  The subscript is the
minimal {\sl word}\/ of the form $a^{n_a} c^{n_c} b^{n_b} d^{n_d}$,
$n_i \geq 0$, matching vector $\mathbf{w}$ under the rule
\begin{equation}\label{notation:rule}
  a^{n_a} c^{n_c} b^{n_b} d^{n_d}
\rightarrow
  ( n_a - n_c ) \mathbf{1} + ( n_b - n_d ) \mathbf{2}
\end{equation}
(obviously, only words of the forms $a^{n_a} b^{n_b}$, $a^{n_a}
d^{n_d}$, $c^{n_c} b^{n_b}$, $c^{n_c} d^{n_d}$ can be minimal.)  For
instance,
\begin{equation}\label{notation:examples}
  G_{ 0,\, X + \mathbf{1} + \mathbf{2} }
\rightarrow
  \zeta_{ab} ,
\qquad
  G_{ P,\, 0 + \mathbf{1} - \mathbf{2} - \mathbf{2} }
\rightarrow
  \xi_{add} ,
\qquad
  G_{ 0,\, 0 - \mathbf{1} }
\rightarrow
  \kappa_c .
\end{equation}

This notation is way more readable than the standard notation, and
allows us to cope with the full form of the density correlator,
including all four terms in (\ref{expansion:twopointfndecomp}).  A C
program was used to generate the full analytic forms in $\zeta_{ab}$
notation for each contribution (tree level and one-loop contributions
I through V), its output being simplified by means of a computer
algebra program.  The simplification thus achieved is dramatic,
witness the results in next section.

\subsubsection{The correlator in $\zeta_{ab}$ notation}

The {\sl full}\/ tree level 2-point topological charge correlator
between 0 and $X$, both in standard and $\zeta_{ab}$ notation, reads:
\begin{equation}\label{formulae:treelevel}
{\setlength\arraycolsep{2pt}
\begin{array}{rcl}
  {\tilde{\mathcal{G}}}^{\mathrm{T}}_{\mathrm{x}}
&=&
{\displaystyle
  2
  \bigg[
    \big(
      G_{0,X+\mathbf{1}+\mathbf{2}}
      + G_{0,X-\mathbf{1}-\mathbf{2}}
      - 2 G_{0,X}
    \big)
    \big(
      G_{0,X+\mathbf{1}-\mathbf{2}}
      + G_{0,X-\mathbf{1}+\mathbf{2}}
      - 2 G_{0,X}
    \big)
}
\\
&&\\
&&
{\displaystyle
\quad
   - \,
    \big(
      G_{0,X+\mathbf{1}} - G_{0,X+\mathbf{2}}
      + G_{0,X-\mathbf{1}} - G_{0,X-\mathbf{2}}
    \big)^2
  \bigg]
}
\\
&&\\
&=&
{\displaystyle
  2
  \bigg[
    ( \zeta_{ab} + \zeta_{cd} - 2 \zeta ) ( \zeta_{ad} + \zeta_{cb} - 2 \zeta )
   -
    ( \zeta_a - \zeta_b + \zeta_c - \zeta_d )^2
  \bigg] .
}
\end{array}
}
\end{equation}

{}From this expression and the asymptotics (\ref{2ndasymp}) and 
(\ref{mixasymp}) of the differences of Green's functions, the leading term 
of the asymptotic behavior for 
${\tilde{\mathcal{G}}}^{\mathrm{T}}_{\mathbf{x}}$ can be easily
computed:
\begin{equation}\label{formulae:asympttreezeta}
  {\tilde{\mathcal{G}}}^{\mathrm{T}}_{\mathrm{x}}
=
  - \,
  \frac{ 2 }{ \pi^2 } \,
  \frac{ 1 }{ |\mathbf{x}|^4 }
 +
  O ( |\mathbf{x}|^{-5} ) .
\end{equation}

The full 1-loop contributions will only be presented in $\zeta_{ab}$
notation (cancelling terms in (III) and (IV) are omitted):

$\bullet$ the graphs with 2-vertices:
\begin{equation}\label{formulae:oneloopone}
{\setlength\arraycolsep{2pt}
\begin{array}{rl}
&
{\displaystyle
  (\mathrm{I_\zeta})
=
  2 \bigg( 1 - \frac{2}{V} \bigg)
}
\\
&\\
&
{\displaystyle
 \times
  \Big[
    ( \zeta_a - \zeta_b + \zeta_c - \zeta_d )
    \sum_P [ ( \xi_a - \xi_b ) ( \eta_{ab} - \eta ) + ( \xi_{ab} - \xi ) ( \eta_a - \eta_b ) ]
}
\\
&\\
&
{\displaystyle
\quad
   - \,
    ( \zeta_{ab} + \zeta_{cd} - 2 \zeta )
    \sum_P ( \xi_a - \xi_b ) ( \eta_a - \eta_b )
   - \,
    ( \zeta_{ad} + \zeta_{cb} - 2 \zeta )
    \sum_P ( \xi_{ab} - \xi ) ( \eta_{ab} - \eta )
  \Big] ,
}
\end{array}
}
\end{equation}

$\bullet$ the X-shaped graphs:
\begin{equation}\label{formulae:onelooptwo}
{\setlength\arraycolsep{2pt}
\begin{array}{rcl}
  (\mathrm{II}_\zeta)
&=&
{\displaystyle
  2 \,
  \sum_P
    [ ( \xi_{ab} - \xi ) ( \xi_{ab} - \xi_{aa} ) - ( \xi_a - \xi_b ) ( \xi_a - \xi_{aab} ) ]
}
\\
&&\\
&&
{\displaystyle
\qquad\quad
\times
    [ ( \eta_{ab} - \eta ) ( \eta_{ab} - \eta_{aa} ) - ( \eta_a - \eta_b ) ( \eta_a - \eta_{aab} ) ]
}
\\
&&\\
&&
{\displaystyle
 + \,
  2 \,
  \sum_P
    [ ( \xi_{ab} - \xi ) ( \xi_{ab} - \xi_{bb} ) + ( \xi_a - \xi_b ) ( \xi_b - \xi_{abb} ) ]
}
\\
&&\\
&&
{\displaystyle
\qquad\quad
\times
    [ ( \eta_{ab} - \eta ) ( \eta_{ab} - \eta_{bb} ) + ( \eta_a - \eta_b ) ( \eta_b - \eta_{abb} ) ] ,
}
\end{array}
}
\end{equation}

$\bullet$ the tadpole graphs:
\begin{equation}\label{formulae:oneloopthree}
{\setlength\arraycolsep{2pt}
\begin{array}{rl}
&
{\displaystyle
  (\mathrm{III}_\zeta)
=
  ( \kappa_a - \kappa )
}
\\
&\\
&
{\displaystyle
 \times
  \Big\{
    ( \zeta_{ab} + \zeta_{cd} - 2 \zeta )
    \sum_P
      \big[
        - ( \xi_a - \xi_b ) ( \eta_{aa} + \eta_{ad} - \eta_{bb} - \eta_{cb} ) 
        - ( \xi_{aa} + \xi_{ad} - \xi_{bb} - \xi_{cb} ) ( \eta_a - \eta_b )
      \big]
}
\\
&\\
&
{\displaystyle
\quad
   +
    ( \zeta_{ad} + \zeta_{cb} - 2 \zeta )
    \sum_P
      \big[
        - ( \xi_{ab} - \xi ) ( \eta_{aab} + \eta_{abb} - \eta_c - \eta_d ) 
        - ( \xi_{aab} + \xi_{abb} - \xi_c - \xi_d ) ( \eta_{ab} - \eta )
      \big]
}
\\
&\\
&
{\displaystyle
\quad
   +
    ( \zeta_a - \zeta_b + \zeta_c - \zeta_d )
    \sum_P
      \big[
         ( \xi_{ab} - \xi ) ( \eta_{aa} + \eta_{ad} - \eta_{bb} - \eta_{cb} )
       + ( \xi_{aa} + \xi_{ad} - \xi_{bb} - \xi_{cb} ) ( \eta_{ab} - \eta )
}
\\
&\\
&
{\displaystyle
\qquad\qquad\qquad\quad\qquad\quad
       + ( \xi_a - \xi_b ) ( \eta_{aab} + \eta_{abb} - \eta_c - \eta_d )
       + ( \xi_{aab} + \xi_{abb} - \xi_c - \xi_d ) ( \eta_a - \eta_b )
      \big]
  \Big\} ,
}
\end{array}
}
\end{equation}

$\bullet$ graphs with six external legs in the FT definition:
\begin{equation}\label{formulae:oneloopfour}
  (\mathrm{IV}_\zeta)
=
  4
  ( \kappa_a + \kappa_{ab} - 2 \kappa )
  \Big[ 
       ( \zeta_{ab} + \zeta_{cd} - 2 \zeta ) ( \zeta_{ad} + \zeta_{cb} - 2 \zeta )
       -
       ( \zeta_a - \zeta_b + \zeta_c - \zeta_d )^2
  \Big] ,
\end{equation}

$\bullet$ the BL difference:
\begin{equation}\label{formulae:oneloopfive}
  (\mathrm{V}_\zeta)
=
  - \,
  4
  ( 2 \kappa_a + \kappa_{ab} - 3 \kappa )
  \Big[ 
       ( \zeta_{ab} + \zeta_{cd} - 2 \zeta ) ( \zeta_{ad} + \zeta_{cb} - 2 \zeta )
       -
       ( \zeta_a - \zeta_b + \zeta_c - \zeta_d )^2
  \Big] .
\end{equation}

Note that (IV$_\zeta$) and (V$_\zeta$) do not involve sums over
lattice points, and that they are proportional to the tree level
expression (\ref{formulae:treelevel}). Their sum reads
\begin{equation}\label{formulae:oneloopfourplusfive}
  (\mathrm{IV}_\zeta) + (\mathrm{V}_\zeta)
=
  - \,
  4
  ( \kappa_a - \kappa )
  \Big[ 
       ( \zeta_{ab} + \zeta_{cd} - 2 \zeta ) ( \zeta_{ad} + \zeta_{cb} - 2 \zeta )
       -
       ( \zeta_a - \zeta_b + \zeta_c - \zeta_d )^2
  \Big] .
\end{equation}

The simplicity of these expressions might be a hint of the existence of a
more direct derivation.

\subsection{Numerical evaluation}
\label{section:numeval}

The procedure we employ to evaluate the correlator numerically follows the 
conventional philosophy adopted in formal perturbation theory: the 
contributions to each order are first evaluated in a finite volume and 
then the thermodynamic limit is taken order by order. It is hoped (though 
far from proven) that by this procedure one obtains an asymptotic 
expansion of the infinite volume correlator. For a more detailed 
discussion of the difficult issues involved see for instance \cite{ps}.

The various perturbative contributions to the charge density correlator
${\tilde{\mathcal{G}}}_{0,X}$ were computed, by means of C programs, for 
$X$ in a finite $400 \times 400$ sublattice of the whole $\mathbb{R}^2$.
\begin{figure}
  \centering
  \psfrag{L}{$L$}
  \psfrag{LI}{$L_{\mathrm{I}}$}
  \psfrag{LO}{$L_{\mathrm{O}}$}
  \includegraphics{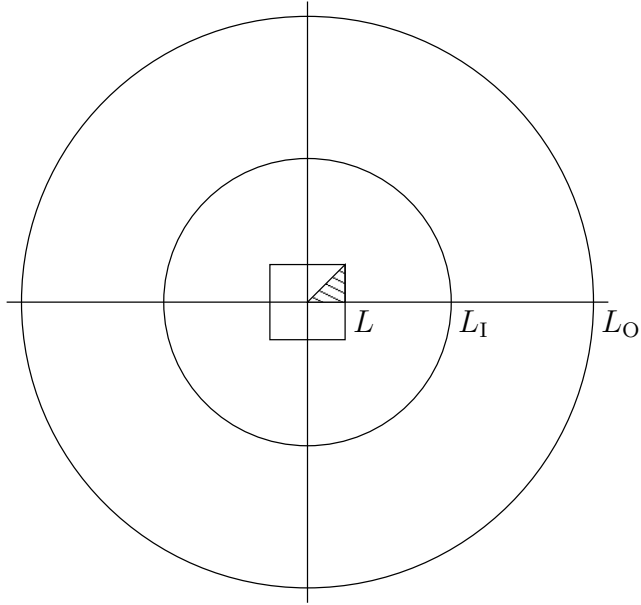}
  \caption{Ranges for the computation: $L$ determines the range of
    $|\mathbf{x}|$ for which topological charge correlators
    $\mathcal{G}_{\mathbf{x}}$ are computed. $L_{\mathrm{O}}$ is a
    cutoff on the sum over Green's functions $G_{\mathbf{x}}$.
    $L_{\mathrm{I}}$ determines the range for which exact values of
    $G_{\mathbf{x}}$, as opposed to asymptotic approximations, are
    used.}
  \label{figure:green}
\end{figure}
Lattice symmetry allows to reconstruct all correlators
in the $2L \times 2L$ square lattice (in our case, $L = 200$) from
those in the shaded region.

As explained in the next sections, the computation of
$\mathcal{G}_{\mathbf{x}}$ involves sums over Green's functions
$G_{\mathbf{x}}$ with $|\mathbf{x}|$ arbitrarily large, and a cutoff
$L_{\mathrm{O}}$ on the summation range is imposed.  Also, since we
have an asymptotic expression for $G_{\mathbf{x}}$, we only need to
compute Green's functions exactly for $|\mathbf{x}|$ in a smaller
region determined by another cutoff $L_I$; for larger $|\mathbf{x}|$
the asymptotic expression is used.

Figure \ref{figure:green} shows the ranges $L$, $L_{\mathrm{I}}$,
$L_{\mathrm{O}}$ involved in the computation.

\subsubsection{Computation of Green's functions}

It is impractical to compute and store an array of all possible values
of (subtracted) Green's functions $G_{\mathbf{x}}$ according to the
exact expression (\ref{green:integralchebyshev}).  It is far more
convenient to store only such values as differ noticeably from the
approximate expression (\ref{green:asymptotics}).  We chose to compute
exact values for $G_{\mathbf{x}}$ with $|\mathbf{x}|$ less than a
certain value $L_{\mathbf{I}}$ (see figure \ref{figure:green}), and
approximate values for the remaining ones as they were needed.

The constant appearing in (\ref{green:asymptotics}) must be computed
for this procedure to be useful.

Working with $L_{\mathrm{I}} = 1000$, exact Green's functions
$G_{(x,y)}$ were computed for the triangular region $0 \leq x \leq
L_{\mathrm{I}}$, $0 \leq y \leq x$.  All those in the region
$|\mathbf{x}| \leq L_{\mathrm{I}}$ can be obtained from them by
symmetry.

Exact values of $L_{\mathrm{I}}$ at the edge $x = L_{\mathrm{I}}$ of
the triangle were used to fit the Green's function space dependence to
(\ref{green:asymptotics}), since this edge is the furthest removed
from the centre and therefore provides the best agreement with the
asymptotic form.

\begin{figure}
  \psfrag{y}{$y$}
  \psfrag{g1000y}{$G_{(1000,y)}$}
  \includegraphics{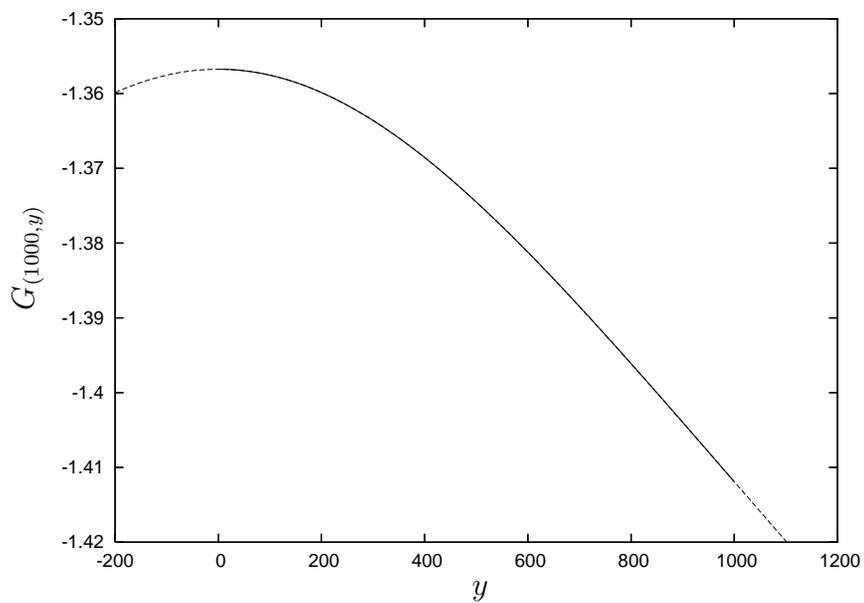}
  \caption{Fit of Green's functions to const - $\frac{1}{2\pi} \ln
    |\mathbf{x}|$ for $\mathbf{x} = (L_i, \, y)$, $0 \leq y \leq
    L_{\mathrm{I}} = 1000$. The dashed line corresponding to the fitted
    dependence is indistinguishable from the computed points in this
    range.}
  \label{fig:greens}
\end{figure}

A value $c = 0.257343$ is obtained for the constant (see figure
\ref{fig:greens}), which agrees with the analytical expression
(\ref{green:valuesofcd}) taken from \cite{shin}.

For $L_{\mathrm{I}} = 1000$, the numerical results for the correlators
are stable with respect to 10\%{} changes in $L_{\mathrm{I}}$.

\subsubsection{Computation of correlators}

The tree level correlator (\ref{formulae:treelevel}), as well as
one-loop contributions (\ref{formulae:oneloopfour}) and
(\ref{formulae:oneloopfive}), do not involve sums over lattice points,
and are readily computed for each $X$.

One-loop contributions (\ref{formulae:oneloopone}),
(\ref{formulae:onelooptwo}) and (\ref{formulae:oneloopthree}) involve
sums over points arbitrarily far away from this sublattice, and a
cut-off must be introduced.  We choose to restrict the sum to points
within the circle of radius $L_{\mathrm{O}}$ about the origin.  At
$L_{\mathrm{O}} = 2000$, results are stable with respect to 10\%{}
changes in $L_{\mathrm{O}}$ or in the shape of the summation domain
(circle vs.~square).  This indicates the absence (or cancellation) of
any infrared divergences connected with the summation over
intermediate lattice points.

Results are presented for the computations outlined in the previous
sections.  Range parameters are $L = 200$, $L_{\mathrm{I}} = 1000$,
$L_{\mathrm{O}} = 2000$.

\begin{figure}
  \psfrag{x}{$x$}
  \psfrag{tgxzero}{${\tilde{\mathcal{G}}}^{\mathrm{T}}_{\mathbf{0},(x,0)}$}
  \psfrag{lgxzero}{${\tilde{\mathcal{G}}}^{\mathrm{L}}_{\mathbf{0},(x,0)}$}
  \includegraphics{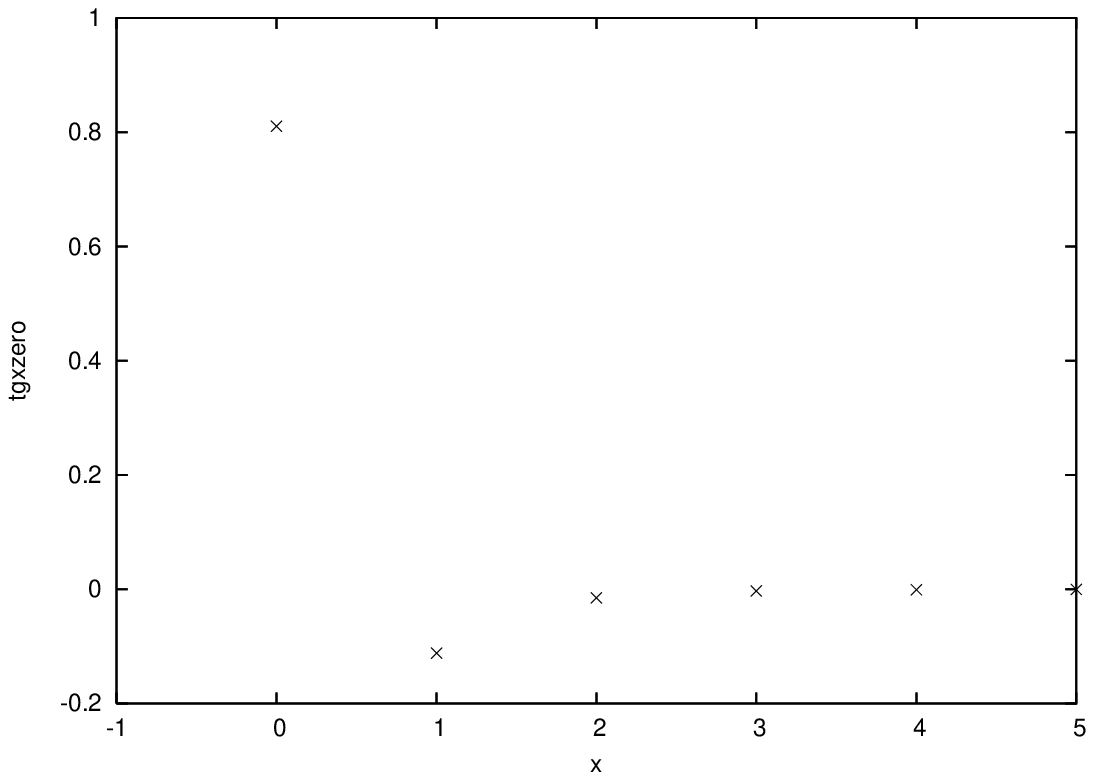}\\
  \includegraphics{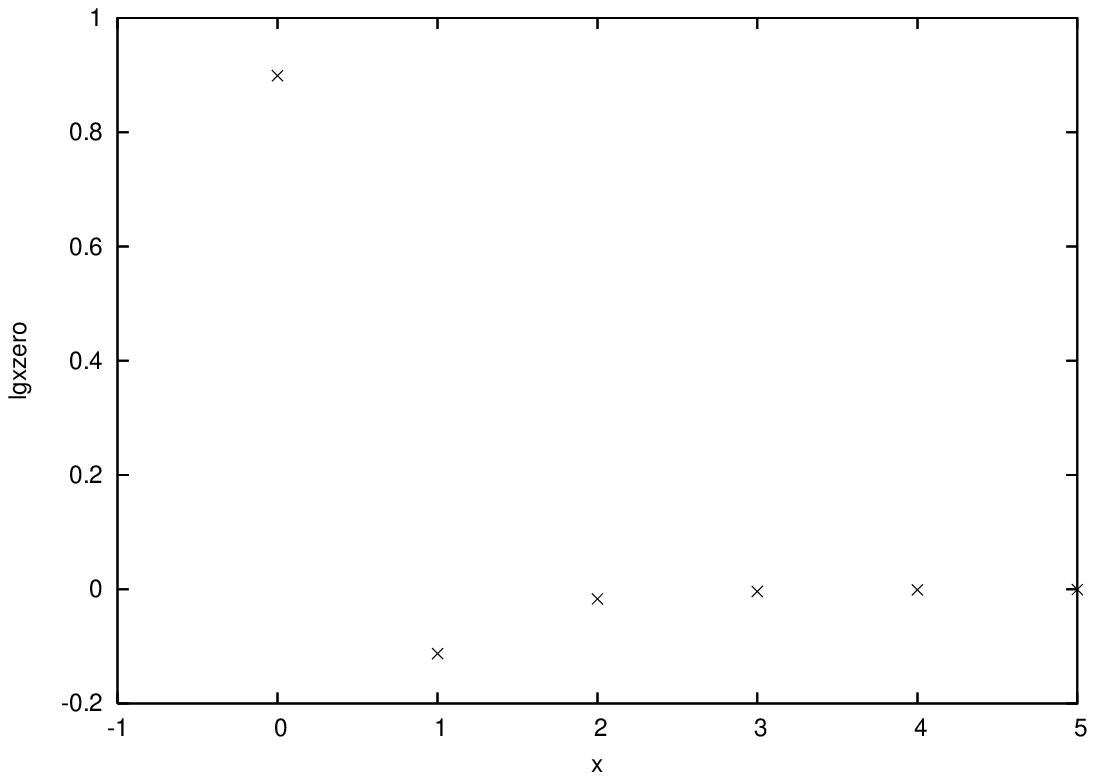}
  \caption{Near-origin behavior of correlator
    ${\tilde{\mathcal{G}}}_{\mathbf{0},(x,0)}$, at tree level and one
    loop, as a function of $x$.}
  \label{fig:nearorigin}
\end{figure}
Figure \ref{fig:nearorigin} shows the (nonuniversal) near-origin tree
level and one loop contributions to the topological charge density
correlator ${\tilde{\mathcal{G}}}_{\mathbf{0},(x,0)}$ as a function of
$x$.  Note that both contributions are positive at the origin,
negative everywhere else, and are suppressed for large $x$.  The peak
at the origin results, in the continuum limit, in the singular contact
term discussed e.g.~in \cite{seilstam}.  The values of the peak at
tree level and one loop level are
\begin{equation}\label{results:peaks}
  {\tilde{\mathcal{G}}}^{\mathrm{T}}_{\mathbf{0}}
\approx
  0.810569 ,
\qquad
  {\tilde{\mathcal{G}}}^{\mathrm{L}}_{\mathbf{0}}
\approx
  0.899007 .
\end{equation}
The tree level peak can also be calculated analytically from
(\ref{green:integralchebyshev}) and (\ref{formulae:treelevel}),
yielding ${\tilde{\mathcal{G}}}^{\mathrm{T}}_{\mathbf{0}} = 8/\pi^2
\approx 0.810570$.

Since this is a perturbative analysis, the topological susceptibility
(\ref{action:susceptibility}) should vanish order by order in
$\beta^{-1}$. Summing the computed correlators over the whole $2 L
\times 2 L$ region, the following results are obtained:
\begin{equation}\label{results:sumtree}
  \sum_{\mathbf{x}} {\tilde{\mathcal{G}}}^{\mathrm{T}}_{\mathbf{x}}
\approx
  1.3 \times 10^{-5} ,
\qquad
  \sum_{\mathbf{x}} {\tilde{\mathcal{G}}}^{\mathrm{L}}_{\mathbf{x}}
\approx
  3.7 \times 10^{-5} ,
\end{equation}
which are compatible with zero and serve as a check of our
computations.  Note, in particular, that the problems deemed to arise
with $\chi_T$ in the continuum limit of the $O(3)$ model do not show
up in this perturbative treatment.

\begin{figure}
  \psfrag{x}{$x$}
  \psfrag{tgx0x4}{$x^4 {\tilde{\mathcal{G}}}^{\mathrm{T}}_{\mathbf{0},(x,0)}$}
  \includegraphics{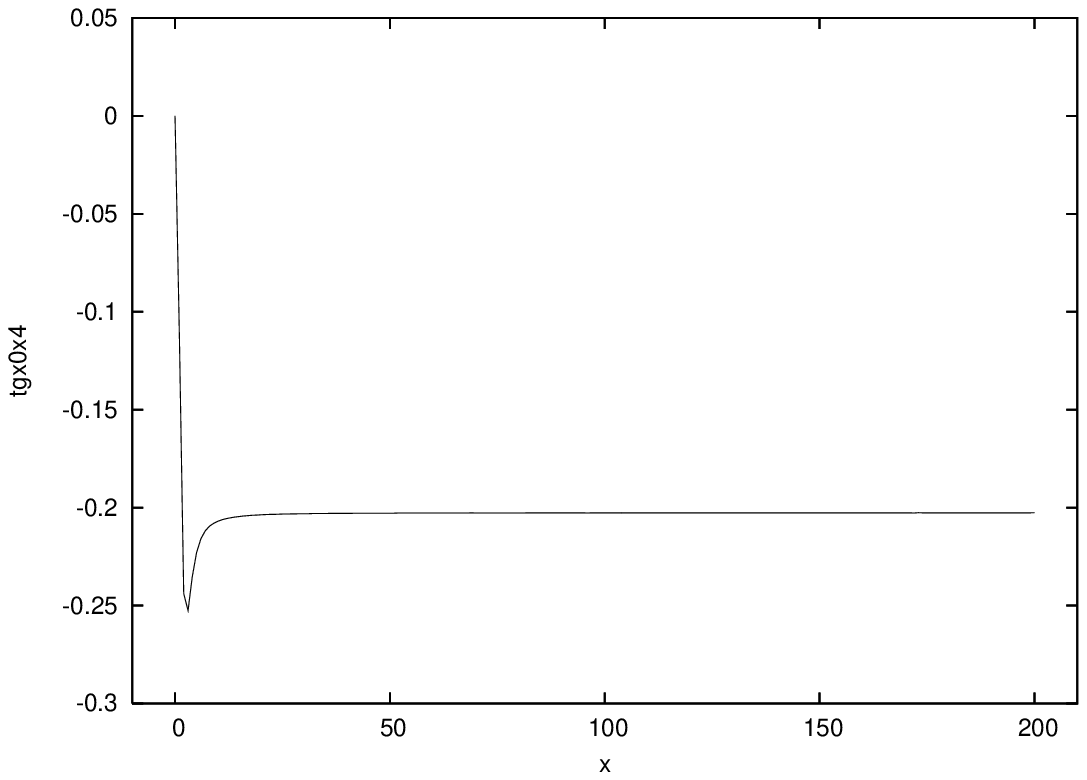}
  \caption{Tree level ${\tilde{\mathcal{G}}}_{\mathbf{0},(x,0)}$
    multiplied by $x^4$ as a function of $x$.}
  \label{fig:tree}
\end{figure}

Figure \ref{fig:tree} shows the tree level contribution multiplied by
$x^4$, as a function of $x$.  This product approaches for large $x$
the constant -0.2026, compatible with (\ref{formulae:asympttreezeta}),
so the tree level result (including all numerical factors) is
\begin{equation}
 {\mathcal{G}}^{\mathrm{T}}_{\mathbf{0},(x,0)}
\sim  -\frac{1} {32 \pi^4 x^4}  
\end{equation}
for $x \gg 1$.

The one-loop contributions (\ref{formulae:oneloopone}) through
(\ref{formulae:oneloopfive}) to the correlator were computed and their
behavior for large $x$ analyzed:
\begin{equation}\label{results:itovasympt}
{\setlength\arraycolsep{2pt}
\left.
\begin{array}{rcl}
  ( \mathrm{I}_\zeta )
&\sim&
{\displaystyle
 - \,
  \frac{ a }{ x^2 } ,
}
\\
&&\\
  ( \mathrm{II}_\zeta )
&\sim&
{\displaystyle
 - \,
  \frac{ b \ln x }{ x^4 } ,
}
\\
&&\\
  ( \mathrm{III}_\zeta )
&\sim&
{\displaystyle
 + \,
  \frac{ a }{ x^2 } ,
}
\\
&&\\
  ( \mathrm{IV}_\zeta + \mathrm{V}_\zeta )
&\sim&
{\displaystyle
 - \,
  \frac{ d }{ x^4 } ,
}
\end{array}\right\}}%
\qquad
  x \gg 1 ,
\end{equation}
with $a=0.101$, $b=0.0653$, $d=0.101$.  These Ans\"atze were obtained
by inspection of the plots and trial and error, except for the last
one, which stems directly from the proportionality of
(\ref{formulae:oneloopfourplusfive}) to (\ref{formulae:treelevel}) and
the fact that $\kappa = 0$, $\kappa_a = -1/4$, which results in the
asymptotics in (\ref{results:itovasympt}) with $d = 1/\pi^2$.

\begin{figure}
  \psfrag{x}{$x$}
  \psfrag{logx}{$\ln x$}
  \psfrag{lgx0x4}{$x^4 {\tilde{\mathcal{G}}}^{\mathrm{L}}_{\mathbf{0},(x,0)}$}
  \includegraphics{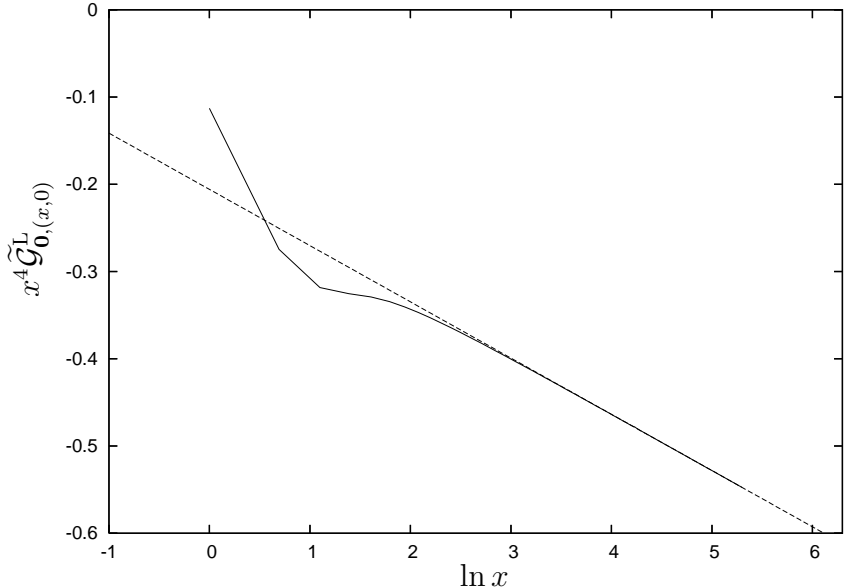}
  \caption{One-loop level ${\tilde{\mathcal{G}}}_{\mathbf{0},(x,0)}$
    multiplied by $x^4$ as a function of $\ln x$, together with a
    linear fit for large $x$.}
  \label{fig:loop}
\end{figure}

The leading terms in (I$_\zeta$) and (III$_\zeta$) cancel, but subleading 
terms $O(x^{-4})$ and $O(x^{-4} \ln x)$ remain. The sum of all one-loop 
contributions (see figure \ref{fig:loop}) behaves as
\begin{equation}\label{results:}
  {\tilde{\mathcal{G}}}^{\mathrm{L}}_{\mathbf{0},(x,0)}    
\sim
 - \,
  \frac{ c_0 + c_1 \ln x }{ x^4 } ,
\qquad
  x \gg 1 ,
\end{equation}
with constants
\begin{equation}
  c_0
=
  2.206 ,
\quad
  c_1
=
  0.0645 .
\end{equation}
Note that these are not identical to $b$ and $d$ in
(\ref{results:itovasympt}), because the subleading terms in the
asymptotic behavior of (I$_\zeta$) and (III$_\zeta$) contribute.

The same analysis was performed along the diagonal of points $(x, \,
x)$ with identical results for the asymptotic forms of the correlator:
full rotational invariance is restored in this regime.

\subsubsection{Renormalization and continuum limit}

We now discuss the continuum limit, again in the spirit of formal
perturbation theory, i.e.~termwise in the perturbative expansion; no
claim can be made that the resulting expansion is asymptotic to a
nonperturbatively defined continuum limit.

Up to one loop we have found for the correlator on the unit lattice
and large lattice distances $|\mathbf{x}|\gg1$
\begin{equation}
\mathcal{G}_{\mathbf{x}}\sim 
- \frac{1}{\beta^2}\  \frac{1}{32\pi^4{|\mathbf{x}}|^4}- 
\frac{1}{\beta^3}\  \frac{1}{64\pi^2{|\mathbf{x}}|^4}(c_0+c_1 \ln 
|\mathbf{x}|)+O\left(\frac{1}{\beta^4}\right)\  . 
\end{equation} 

To obtain a continuum limit we have to introduce the lattice spacing
$a$, rescale $x=y/a$ and make $\beta$ dependent on the cutoff $a$. We
also have to rescale the correlator $\mathcal{G}$ by a factor $a^4$
according to its engineering dimension, to obtain the correlator
$\mathcal{G}_{\mathbf{y}}^a$ in continuum normalization
\begin{equation}\label{contnormcorr}
\mathcal{G}_{\mathbf{y}}^a = a^4\mathcal{G}_{\mathbf{x}a}
\sim - \frac{1}{\beta(a)^2}\  \frac{1}{32\pi^4{|\mathbf{y}}|^4}-
\frac{1}{\beta(a)^3}\  \frac{1}{64\pi^2{|\mathbf{y}}|^4}\left(c_0+c_1 \ln
\frac{|\mathbf{y}|}{a}\right)+O\left(\frac{1}{\beta(a)^4}\right)\ ,
\end{equation}
valid asymptotically for $|\mathbf{y}|\gg a$.

A continuum limit should exist if we let $\beta$ depend logaritmically
on $a$ according to the one-loop Callan-Symanzik $\beta$ function for
the $O(3)$ model \cite{bz}:
\begin{equation}
  \frac{1}{\beta(a)}
=
  \frac{1}{\beta_0-\frac{1}{2\pi} \ln(\mu a)} + O ( \beta_0^{-3} )
=
  \frac{1}{\beta_0}
  \left( 1 + \frac{1}{2\pi\beta_0} \ln(\mu a) + O ( \beta_0^{-2} ) \right)\ .
\end{equation}
Inserting this in (\ref{contnormcorr}) and reexpanding to order
$\beta_0^{-3}$ we see that the terms proportional to $\ln a$ cancel if
$c_1 = 2 / \pi^3$, which is consistent with the value 0.0645
produced by our numerical computation; the continuum limit of the
correlator is then
\begin{equation}\label{contcorrmu}
  \lim_{a\to 0} \mathcal{G}_{\mathbf{y}}^a
\equiv
  \mathcal{G}_{\mathbf{y}}
=
  - \frac{1}{\beta_0^2}\ \frac{1}{32\pi^4 |{\mathbf{y}}|^4}
  \left( 1 +
    \frac{1}{\pi \beta_0}
    \ {\ln \left[ |\mathbf{y}| \, \mu \, \mathrm{e}^{\frac{\pi^3 c_0}{2}} \right]}
  \right)
  + O \left( \beta_0^{-4} \right)\ .
\end{equation}
By choosing a renormalization scale $\mu = \mu_0\mathrm{e}^{-\,\frac{\pi^3
c_0}{2}}$ we obtain
\begin{equation}\label{contcorr}
  \lim_{a\to 0} \mathcal{G}_{\mathbf{y}}^a
\equiv
  \mathcal{G}_{\mathbf{y}}
=
  - \frac{1}{\beta_0^2}\ \frac{1}{32\pi^4 |{\mathbf{y}}|^4}
  \left( 1 +
    \frac{1}{\pi \beta_0}
    \ {\ln(\mu_0 |\mathbf{y}|) }
  \right)
  + O \left( \beta_0^{-4} \right)\ .
\end{equation}
If we had chosen the FT instead of the BL definition, the only
difference wuld have been a change in $c_0$, corresponding to
different renormalization scale, so to the order considered the two
definitions are related by a finite renormalization.

Equation (\ref{contcorr}) is now valid for all $\mathbf{y}$, since in
the limit $a\to 0$ only the asymptotic behavior of
$\mathcal{G}_{\mathbf{x}}$ survives.

We can compare this result with the Renormalization Group improved tree 
level result \cite{bn,vic} 
\begin{equation}\label{rgtree}
\mathcal{G}_{\mathbf{y}}= - \frac{1}{32\pi^4\beta({\mathbf{y}})^2}\ 
\frac{1}{|\mathbf{y}|^4}\ ,
\end{equation}
if we reexpand (\ref{rgtree}) to order $\beta_0^{-3}$, using the one-loop 
RG flow
\begin{equation}
\beta(\mathbf{y})=\beta(\mathbf{y}_0)- 
\frac{1}{2\pi} \ln \frac{ |\mathbf{y}| }{ |\mathbf{y}_0| }\ .  
\end{equation}
The expressions (\ref{contcorr}) and (\ref{rgtree}) agree to order
$\beta_0^{-3}= \beta(\mathbf{y}_0)^{-3}$ if the renormalization scales
are chosen appropriately.

So we have established that our one-loop calculation supports the
softening of the short distance behavior predicted by the RG improved
tree level result.

\section{Conclusions}
We computed the two-point functions of the topological charge density
perturbatively to one loop and found consistency with the RG improved
tree level perturbative result, indicating a softening of the short
distance singularity compared to the naive tree level result. This
means that at the level of formal one-loop perturbation theory the
requirements of the two positivities analyzed in \cite{clash} are
indeed satisfied.

But we should point out again that a mathematical justification of the
procedures of formal perturbation theory (interchange of the weak
coupling limit with the thermodynamic and continuum limits) does not
exist, therefore it would be very interesting to see if this softening
is really present at the nonperturbative level.  Numerical checks
using Monte Carlo simulations are presumably very difficult, however,
since the check requires identifying logarithmic corrections to a
rather large power.

\section*{Acknowledgements}
We are grateful to Peter Weisz for a critical reading of the
manuscript and for drawing our attention to reference \cite{shin}.


\end{document}